\documentclass{article}

\usepackage{arxiv}

\usepackage[utf8]{inputenc} 
\usepackage[T1]{fontenc}    
\usepackage{hyperref}       
\usepackage{url}            
\usepackage{booktabs}       
\usepackage{amsfonts}       
\usepackage{nicefrac}       
\usepackage{microtype}      
\usepackage{lipsum}		
\usepackage{graphicx}
\usepackage{doi}
\usepackage{amsmath}

\title{Predictions of the neutrino oscillations parameters}

\author{ \href{https://orcid.org/0000-0002-2509-5048}{\includegraphics[scale=0.06]{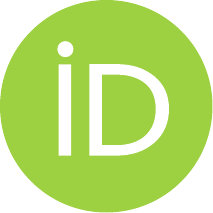}\hspace{1mm}Ivan Arraut}\\
	 Institute of Data Engineering and Science (IDEAS)\\
 and Institute of Science and Environment,\\
 University of Saint Joseph, \\
 Estrada Marginal da Ilha Verde, 14-17, Macao, China\\
	\texttt{ivan.arraut@usj.edu.mo} \\
}

\renewcommand{\shorttitle}{\textit{arXiv} Template}

\hypersetup{
pdftitle={A template for the arxiv style},
pdfsubject={q-bio.NC, q-bio.QM},
pdfauthor={Ivan Arraut},
pdfkeywords={Dark Energy, Zero point quantum fluctuations, Cosmological constant},
}

\begin{document}
\maketitle

\begin{abstract}
We demonstrate that the observed parameters, related to the flavor oscillation of the neutrinos, are consistent with a system with three degenerate ground states. These ground states must have the same energy and they are connected by a symmetry with its corresponding generator. This symmetry is not spontaneously broken because the neutrino never selects a specific vacuum state. Instead, the neutrino keeps oscillating between the three ground states as it moves through spacetime. The order parameter in this case is the neutrino flavor and then the superposition of all the possible vacuum flavor states for one neutrino gives a trivial (zero) result. From this trivial condition, we can find constraints over the observed mixing angles $\theta_{ij}$. By using these constraints we make the predictions of the mixing angles which are in agreement with the observations. Subsequently, complementing these arguments with a triangular formulation of the neutrino oscillation, we find relations between the the mass eigenvalues and the mixing angles, making again predictions consistent with the observations and consistent with a normal order in hierarchy with $m_3>>m_2\approx m_1$. Finally, we analyze the symmetry related to the flavor oscillation.
\end{abstract}


\section{Introduction}

Neutrinos were initially postulated as a necessity of maintaining the energy-momentum conservation in some processes involving the weak-interaction \cite{the_neutrino}. These particles are fermions coming in three different flavors, corresponding to electronic, muonic and tau \cite{pdg}. When Weinberg, Salam and Glashow formulated the electroweak theory, they introduced the neutrino as a massless particle \cite{glashow, salam, weinberg}. In fact, inside this theory, the neutrinos are left-handed and massless \cite{zeromass} and the Higgs cannot generate their mass dynamically. In this way, the mechanism for generating the neutrino mass is not well-understood. Even being massive, the neutrino mass is extremely small. The absence of right-handed neutrinos does not allow a description in terms of the two Weyl components of fermions in quantum field theory. Hence, the usual mass term in the Lagrangian cannot be built, and it is set to zero manually in the Standard Model \cite{weyl}. The discovery of the massive character of the neutrinos came after it was understood that they oscillate between different flavors after being generated from a weak-process and coming initially in a single flavor \cite{sk,sno}. The massive character of the neutrinos can be also elucidated through their flavor oscillation \cite{pontecorvo,mns}. There are three known flavors for the neutrinos and three different eigenvalues for the masses. The mass eigenvalues do not correspond to the flavor eigenvalues. In this way, one neutrino with some specific flavor has an effective mass which is a linear combination of three different masses labeled as $m_1$, $m_2$ and $m_3$ and vice-versa, one neutrino with some specific mass eigenvalue can be seen as a mix of three different flavors with their corresponding masses. We can convert from the flavor basis to the eigenvalue basis by only calculating the inverse of the neutrino mixing matrix. The experiments usually measure the neutrino mixing angles and the square of the mass differences $\vert\Delta m_{ij}^2\vert=\vert m_j^2-m_i^2\vert$ \cite{parameters}. The experiments still do not make any specific measure of the individual values of the neutrino masses $m_i$. The experiments cannot determine whether there is a Normal Ordering (NO) such that $m_3>>m_2\approx m_1$, or an Inverted Order (IO) such that $m_1\approx m_2>>m_3$. Several theoretical proposals have been done which intend to explain the mechanism behind the neutrino oscillation. The seesaw mechanism for example, introduces a heavy neutrino, besides the light neutrino with the corresponding flavors. This mechanism can in principle explain the origin of the small mass of the neutrino via symmetry arguments \cite{Seesaw, Seesaw2}. Discrete non-Abelian symmetries have been also introduced in order to explain the origin of the neutrino mass \cite{Delta1, Delta2}. The related tribimaximal mixing models for certain symmetries were introduced in \cite{Delta2, Delta3, Delta4}. The tribimaximal models on their original versions were ruled out by experiments. However, some extended versions of the model are still considered in the literature. In 2024 the author of this paper formulated a triangular model suggesting constraints between the mixing angles and the mass eigenvalues \cite{Triangle1, Triangle12}. However, by that time a clear formulation in terms of symmetries was not understood. It was not understood neither the role of the flavor states as an order parameter of the system. In 2025, a year later, another author used the same triangle formulation \cite{Triangle2}, but in this case there was no intention to connect the triangular formulation with symmetry arguments. Additionally, although the location of the mixing angles in both proposals is the same, the interpretation of the meaning of the sides of the triangles is different. It seems that the author in \cite{Triangle2} was not aware about the results proposed in \cite{Triangle1, Triangle12}. In this paper we explain how the flavor oscillation of a single neutrino is equivalent to a periodic change of the vacuum state perceived by the same particle. We can then interpret the neutrino flavor as an order parameter. In this way, the superposition of all the vacuum states defined by the different flavors gives a trivial (zero) result, which generates important constraints between the mixing angles. Interestingly, when we change the basis from the flavor basis to the eigenvalue basis, the angle $\theta_{13}$ is fixed, while $\theta_{12}$ and $\theta_{23}$ exchange roles. Regarding the values of mass, we generate an extension to the geometrical model proposed in \cite{Triangle1, Triangle12} in order to make some predictions on the possible values of mass $m_1$, $m_2$ and $m_3$. We implement the Quantum Yang Baxter Equations (QYBE), previously understood as the most general way to understand the Nambu-Goldstone theorem \cite{M1, M2, M3}, such that the interpretation of the flavor states as the order parameter emerge naturally. The QYBE also suggest that the signature of the mixing angle is not relevant and that rather its absolute value is the important one. This result will become relevant when we obtain the mixing angles from the restrictions emerging from the vacuum flavor states taken as the order parameter of the system. Finally, we analyze the potential symmetry behind the oscillation process, together with its charge. This is not a trivial task and in this paper we just give a preliminary analysis looking for this purpose. We also try to interpret the symmetry patterns in connection with the known and related electroweak symmetries, giving again a preliminary taste of what will be the final theory. The paper is organized as follows: In Sec. (\ref{mix}), we describe the standard neutrino oscillation pattern. In Sec. (\ref{Vacuumpers}), we interpret the same phenomena from the perspective of three different vacuum states. In Sec. (\ref{Deeper}), we take a look into the neutrino oscillation and the potential symmetry involved in the process. Here we interpret the approximation $m_1\approx m_2$ in terms of the mixing angle $\theta_{13}$. In Sec. (\ref{YANG}), we use the approximate triangular relation between neutrino flavors, in order to make predictions on the mass eigenvalues. In the same section, we make predictions over the neutrino mixing angles by using the fact that the neutrino flavor is an order parameter of the system. In Sec. (\ref{Predictionsanalysis}), we make a discussion about future analysis, predictions and potential experiments for testing the results exposed in this paper. Finally, in Sec. (\ref{concl}) we conclude.

\section{The neutrino mixing matrix}   \label{mix}

The neutrino mixing matrix represents the mass eigenvalues $m_i$ as a linear combination of the flavor eigenvalues $m_\alpha$ \cite{pontecorvo,mns}. The state of a neutrino can, in general, be expressed as

\begin{equation}
\nu_\alpha=\sum_iU_{\alpha i}\nu_i.
\end{equation}
Here the subindex $\alpha$ corresponds to some specific flavor and the subindex $i$ corresponds to the mass eigenvalues. In the same way as we can express the mass eigenvalues as a superposition of flavor eigenvalues, the opposite is also true, namely, the flavor eigenvalues can be expressed as the superposition of mass eigenvalues. The matrix $U_{i \alpha}$ for the case where the experiments just look for two flavors, is defined as 

\begin{equation}
U=\begin{pmatrix}
cos\theta & sin\theta\\
-sin\theta & sin\theta 
\end{pmatrix}, 
\end{equation}
This matrix is just a unitary rotation in two dimensions. The matrix is in a basis which allows us to express each neutrino flavor as a linear combination of the mass eigenvalues. In the most general situation in the standard three neutrino picture (without any other neutrino species), we have three different types of neutrinos. In such a case, the flavors and the mass eigenvalues are related through a three-dimensional matrix, which represents a rotation in a three dimensional space. The matrix, in the flavor basis is given by 

\begin{equation}   \label{matrix}
\begin{bmatrix}
c_{12}c_{13} & s_{12}c_{13} & s_{13}\\
-s_{12}c_{23}-c_{12}s_{23}s_{13} & c_{12}c_{23}-s_{12}s_{23}s_{13} & s_{23}c_{13}\\
s_{12}s_{23}-c_{12}c_{23}s_{13} & -c_{12}s_{23}-s_{12}c_{23}s_{13} & c_{23}c_{13}
\end{bmatrix}.
\end{equation}
Here $c_{ij}=cos(\theta_{ij})$ and $s_{ij}=sin(\theta_{ij})$. This matrix is evidently a unitary three-dimensional rotation matrix but instead of being represented through the Euler angles, it is represented as a function of the Tait-Bryan angles \cite{TB}. The matrix is an operator related to a simultaneous rotation through the axis representing the electronic, the muon and the tau flavors. We must remark that the matrix (\ref{matrix}) represents the situation where the CP symmetry is not violated. If the CP symmetry is violated, an additional parameter emerges as it can be seen in \cite{pontecorvo,mns}. In this letter however, we will focus on the case where $CP$ violations are ignored and then we do not include the related parameter inside the matrix (\ref{matrix}).
\section{Neutrino oscillation: Vacuum perspective}   \label{Vacuumpers}

If we analyze the neutrino oscillations from the perspective of the vacuum state, then we can suggest three different ground states corresponding to the electron neutrino, tau neutrino and muon neutrino. Then a single neutrino with a mass eigenvalue $m_i$, will perceive the oscillation of its ground state, moving from one state where the neutrino is a $\tau$-neutrino, towards another one where the neutrino is a $\mu$-neutrino and subsequent $e^-$-neutrino. Then the vacuum state perceived by the neutrino is in reality a linear superposition of the three ground states $\vert0>_{e^-}$, $\vert0>_\mu$ and $\vert0>_\tau$. This is evident if we evaluate the inverse of the mix matrix (\ref{matrix}). We can define some creation and annihilation operators for each one of the Hilbert spaces containing the corresponding flavors. Then when the vacuum state is such that the neutrino is $e^-$-neutrino, we can define the creation and annihilation operators as

\begin{equation}
\hat{a}^+_e\vert0>_e=\vert\nu>_e,\;\;\;\hat{a}_e\vert0>_e=0.
\end{equation}
However, the vacuum perceived by the neutrino with a mass $m_i$ is an oscillatory vacuum, defined more properly as 

\begin{equation}   \label{standardvac}
\hat{a}_i\vert0>_i=0.    
\end{equation}
We can relate the vacuum $\vert0>_i$ with the flavor vacuum $\vert0>_\alpha$ through the relation

\begin{equation}   \label{super}
\vert0>_i=\sum_\alpha a_\alpha \vert0>_\alpha,
\end{equation}
which is actually connected with the mixing matrix (\ref{matrix}). In eq. (\ref{super}), the coefficients $a_\alpha$ represent the factors containing the mixing angles $\theta_{ij}$. In fact, eq. (\ref{super}), can be expressed as 

\begin{eqnarray} \label{linear_comb}
\vert0 \rangle_e & = & c_{12}c_{13}\vert0\rangle\rangle_1+s_{12}c_{13}\vert0\rangle\rangle_2+s_{13}\vert0\rangle\rangle_3, \nonumber \\
\vert0 \rangle_\mu & = & (-s_{12}c_{23}-c_{12}s_{23}s_{13})\vert0\rangle_1 + \nonumber \\
& & (c_{12}c_{23}-s_{12}s_{23}s_{13})\vert0\rangle_2+s_{23}c_{13}\vert0\rangle_3, \nonumber \\
\vert0 \rangle_\tau & = & (s_{12}s_{23}-c_{12}c_{23}s_{13})\vert0\rangle_1 + \nonumber \\
& & (-c_{12}s_{23}-s_{12}c_{23}s_{13})\vert0\rangle_2+c_{23}c_{13}\vert0\rangle_3.
\end{eqnarray}
When the neutrino travels, it is permanently changing from flavor because it is oscillating between the three different vacuum states defined in eq. (\ref{linear_comb}). The flavor vacuums are degenerate because when the neutrino oscillates, it still maintains the same energy. The neutrino never selects any of these vacuums, perhaps because the external interactions are very weak in order to force the neutrino to select any of these ground states. In fact, the neutrino only selects one specific vacuum at the moment when it interacts with the detectors. The standard approach to the neutrino oscillation problem, suggests that flavor states are one-particle states. However, the particle related to the state must emerge from some vacuum state. Then here we are taking the vacuum state for this particle and we consider the neutrino flavor as the order parameter of the system. In order to prove that the flavor is the order parameter, there is a consistency condition to satisfy if we interpret the three different vacuum flavors in eq. (\ref{linear_comb}) as the three degenerate vacuum states. The consistency condition suggests that if we sum over all possible flavors, or equivalently, over all the possible vacuum states representing the flavors, we then recover the trivial vacuum state. This means that the flavor inside the neutrino oscillation process, must satisfy the condition

\begin{equation}    \label{SumofVacumms}
\vert0 \rangle_e+\vert0 \rangle_\mu+\vert0 \rangle_\tau=0.
\end{equation}
If we use the relations defined in eq. (\ref{linear_comb}), then we get from eq. (\ref{SumofVacumms})

\begin{eqnarray}   \label{ImportantConditions}
c_{12}c_{13}-s_{12}c_{23}-c_{12}s_{23}s_{13}+s_{12}s_{23}-c_{12}c_{23}s_{13}=0,\nonumber\\
s_{12}c_{13}+c_{12}c_{23}-s_{12}s_{23}s_{13}-c_{12}s_{23}-s_{12}c_{23}s_{13}=0,\nonumber\\
s_{13}+s_{23}c_{13}+c_{23}c_{13}=0.\;\;\;
\end{eqnarray}
The emergence of this condition can be understood better if we see that the condition (\ref{SumofVacumms}) has to be applied to each component of eq. (\ref{linear_comb}). Then for example, the first equation in (\ref{ImportantConditions}), emerges from the sum over the $\vert0>_1$-components in eq. (\ref{linear_comb}) after considering the condition (\ref{SumofVacumms}). Similar analysis for the other two equations, which correspond to the components $\vert0>_2$ and $\vert0>_3$. The first and the second equations in (\ref{ImportantConditions}) are redundant. They represent the same equation but shifted by an angle of $-\frac{\pi}{2}$ over the angle $\theta_{12}$. In fact, the second equation inside (\ref{ImportantConditions}), can be expressed as 

\begin{equation}
\bar{c}_{12}c_{13}-\bar{s}_{12}c_{23}-\bar{c}_{12}s_{23}s_{13}+\bar{s}_{12}s_{23}-\bar{c}_{12}c_{23}s_{13}=0.    
\end{equation}
In this equation, we have defined $\bar{c}_{12}=cos\left(\theta_{12}-\frac{\pi}{2}\right)$ and $\bar{s}_{12}=sin\left(\theta_{12}-\frac{\pi}{2}\right)$.
This result is identical to the first equation in (\ref{ImportantConditions}), but considering this time the shift in the angle $\theta_{12}$ by $\frac{\pi}{2}$. This shift will not affect the experimental outcomes because normally the experiments detect the mixing angles through the relation $sin^2(2\theta_{12})=sin^2(2\theta_{12}-\pi)$. This means that the first and the second equation inside the relations (\ref{ImportantConditions}) are redundant. Another way to see this redundancy, is by noticing that the first and second equations in (\ref{ImportantConditions}) are related by the derivative with respect to $\theta_{12}$. Then if we take the derivative with respect to $\theta_{12}$ of the second equation, we recover the first one. Then both equations will give us redundant angles, namely, angles which bring out the same observables. From the third equation inside (\ref{ImportantConditions}), we can solve it if we input one of the observed values for the mixing angles. If we divide this equation by $c_{13}$, we get

\begin{equation}   \label{1st}
tan(\theta_{13})=-sin(\theta_{23})-cos(\theta_{23}).
\end{equation}
If we input the experimentally observed mixing angle $\theta_{13}=\pm 8.47 deg$, this angle corresponds to the observation $sin^2(2\theta_{13})=0.085$. Before proceeding, just for the sake of simplicity, we will map all the angles as $\theta_{ij}\to-\theta_{ij}$. In such a case, the equation (\ref{1st}), becomes

\begin{equation}   \label{1st1}
tan(\theta_{13})=-sin(\theta_{23})+cos(\theta_{23}).  
\end{equation}
Then if we input now the value $\theta_{13}=8.47deg$, then we get $\theta_{23}\approx 39deg$. In this case, we are considering the smallest angle (in absolute value), consistent with the trigonometric relation. For this reason, the angle $\theta_{23}$ is taken over the first quadrant. In order to find $\theta_{12}$, we have to use one of the equations available in (\ref{ImportantConditions}). However, before doing this, we will also exchange the sign of the angles in the same way as we did with the third equation within the same expressions. In this way, we obtain the following

\begin{eqnarray}   \label{ImportantConditions2}
c_{12}c_{13}+s_{12}c_{23}-c_{12}s_{23}s_{13}+s_{12}s_{23}+c_{12}c_{23}s_{13}=0,\nonumber\\
-s_{12}c_{13}+c_{12}c_{23}+s_{12}s_{23}s_{13}+c_{12}s_{23}-s_{12}c_{23}s_{13}=0,
\end{eqnarray}
If we input $\theta_{13}$ and $\theta_{23}$ inside these equations, we get two results which are both consistent with the observations. The results are $\theta_{12}\approx 54deg$ or $\theta_{12}\approx -36deg$. Both angles are shifted by $90deg$, reproducing the same observables experimentally. Most of the values obtained for the angles are inside the range of values observed experimentally. However, for the angle $\theta_{12}\approx 54 deg$, the situation is complex but interesting. This angle reproduces the right value for $sin^2(2\theta_{12})$ but the angle predicted by the experiments is $\theta_{12}\approx 36 deg$, which is precisely the complementary angle to $\theta_{12}=54 deg$. We can then conclude that the reason for this result to appear in this way, is the conservation of probability under the exchange of signatures of the mixing angles \cite{parameters}. In fact, the present results show that while two of the angles, namely, $\theta_{13}$ and $\theta_{23}$ correspond to the usual observations, the order parameter condition (\ref{ImportantConditions2}) suggest that the third angle $\theta_{12}$ must take the complementary value $\theta_{12}\pm\pi/2$. This guarantees that the probability of transition between flavors does not change under the variation of signature of the angle $\theta_{ij}$ \cite{parameters}. For this reason, in this paper we are taking the mixing angles in agreement with their absolute values $\vert\theta_{ij}\vert$, particularly when we deal with the neutrino flavor as an order parameter.

\section{A deeper look into the vacuum symmetry for the neutrino oscillation}   \label{Deeper}

In this paper we suggest that the neutrinos live over three possible vacuum states, determined by the flavor states of the system. In this way, although the neutrino does not break the symmetry associated with the oscillation spontaneously, it is still important to illustrate here the mechanism of spontaneous symmetry breaking and then compare it with what the neutrino is experiencing. Here we will analyze the standard mechanism of spontaneous symmetry breaking \cite{M1, M2, M3, Nambu2, Nambu3, Nambu4}, as well as the fact that a well-defined order parameter will have a trivial result when it is summed over all its possible values. The typical potential where the mechanism of spontaneous symmetry breaking emerges, is the "Mexican" hat appearing on the figure (\ref{fig2}). In this case, we have a degenerate ground state plus a central unstable equilibrium state. In general, the mechanism of spontaneous symmetry breaking occurs when the system, located initially over the unstable equilibrium state, will be looking for stability by moving towards a new equilibrium position. In this initial condition, the order parameter of the system satisfies $<0\vert\phi_1\vert0>=0$. 
There are several possible new vacuum states but the system selects the one pointing on the direction of the external perturbation. The physical character of the external perturbation depends on the system under analysis. Then for example, in a ferromagnet, the external perturbation is the external magnetic field. Once the small perturbation touches the system over the central unstable equilibrium, then a final vacuum state will be selected. Once the system reaches its new equilibrium condition, breaking then the corresponding symmetry spontaneously, the condition over the order parameter suggests that 

\begin{equation}   \label{Situation}
<0\vert\phi_1\vert0>\neq0.    
\end{equation}
This inequality emerges because now the system has selected one of the degenerate ground states. Yet still, if we sum all the possibilities for the order parameter, or equivalently, if we sum over all the possible vacuum states breaking a symmetry spontaneously, then we get a trivial result, recovering in this way the initial ground state. One way to represent this important result is by using the Quantum Yang Baxter Equations, in the form proposed in \cite{M1, M2, M3}. 
Here we will take it as the order parameter with its flavor as a label. Inside the electroweak theory, it is normal to see some flavor change happening in certain processes involving quarks and the exchange of the W-bosons \cite{zeromass}. Still, the quarks at the end of the flavor exchange processes have a well-defined flavor. Among the processes involving flavor exchange for quarks we have the beta-decay, where the d quark converts into the u quark. This can be seen from the neutron decay, described as \cite{Klad}

\begin{equation}   \label{process1la}
n\to p+e^-+\bar{\nu}_e.    
\end{equation}
From the perspective of the quarks inside the baryons, after ignoring any effects related to the strong interaction, we can see that the process (\ref{process1la}) is equivalent to \cite{Klad}

\begin{equation}   \label{process2la}
d\to u+e^-+\bar{\nu}_e.    
\end{equation}
This is true since under the weak-interaction, the other two quarks inside the neutron just act as spectators. It is important to remark that the process (\ref{process2la}) involves the exchange of a $W$-bosons, which is necessary in order to guarantee charge conservation. The hadronic processes inside the weak interaction are in general interesting because they contain in some cases flavor changes. At the same time, they involve the emergence of a neutrino (or antineutrino). This is an important detail to consider in future in order to understand the mechanism behind the neutrino mass. Then for example, for the case of decaying mesons, we have the case where

\begin{equation}   \label{HBR}
\pi^-\to \mu^-+\bar{\nu}_\mu,     
\end{equation}
with the same decaying for $\pi^+$ but with the corresponding exchange of signs. While the process (\ref{HBR}) has a very high branching ratio, the corresponding decaying involving electrons has a very low branching ratio due to the well-known helicity constraints \cite{Klad}. Yet still, the process with small branching ratio occurs and both, the muonic and the electronic process implies the interaction of two different flavors of quarks with the corresponding exchange of the W-boson. An explicit flavor change occurs for the process

\begin{equation}
\pi^-\to\pi^0+e^-+\bar{\nu}_e.    
\end{equation}
In this case, this process also involves the exchange of a W-boson, with the corresponding exchange of the quarks $d\to u$ and $\bar{u}\to\bar{d}$ \cite{Klad}. Other more complex processes, involving the strange quark (s) for example, implies some flavor mixing for the final quarks. This mixing, proposed originally by Cabibbo, together with the GIM mechanism which avoids flavor changes for neutral currents, is complemented with the work done by Kobayashi and Maskawa \cite{Klad}.
The neutrino's behavior on the other hand is outside the standard model. Its flavor change occurs after certain processes are supposed to finish and it keeps oscillating among the different flavors. From the perspective of this paper, this means that it is unable to select any of the possible final ground states. The oscillation pattern can then be perceived through the algebra of the matrices containing the mixing angles. For example, we can define the mixing matrix (\ref{matrix}) as the successive application of the operations

\begin{equation}
\Theta_{23}\Theta_{13}\Theta_{12},    
\end{equation}
where the operators $\Theta_{ij}$ are defined as

\begin{eqnarray}
\Theta_{12}=  \label{matrix}
\begin{bmatrix}
c_{12} & s_{12} & 0\\
-s_{12} & c_{12} & 0\\
0 & 0 & 1
\end{bmatrix}, \;\;\;\;
\Theta_{23}=\begin{bmatrix}
1 & 0 & 0\\
0 & c_{23} & s_{23}\\
0 & -s_{23} & c_{23}
\end{bmatrix},\;\;\;\;
\Theta_{13}=\begin{bmatrix}
c_{13} & 0 & s_{13}\\
0 & 1 & 0\\
-s_{13} & 0 & c_{13}
\end{bmatrix}.
\end{eqnarray}
We can construct an algebra from these matrices, by taking into account that the neutrino flavor is the order parameter of the system. This algebra however, is just the same algebra of the $SO(3)$ matrices which is non-trivially homomorphic to $SU(2)$ \cite{Homomorphism}. Before discussing the possible symmetries involved with the flavor oscillation, we will illustrate two possible methods for predicting the mass eigenvalues $m_1$, $m_2$ and $m_3$. First we will try a geometric method based on triangular relations and subsequently, we will compare the results with the argument taking the neutrino as an order parameter. Any discrepancy between the results obtained by using the two methods will be interpreted.

\subsection{The hidden symmetry}

The symmetry governing the dynamics of the neutrinos, is a symmetry related to the possible flavors taken by the neutrino. Here we analyze the group structure of this symmetry. The symmetry which the neutrinos cannot break when they oscillate among the three different states, will be taken to be $SU(3)_\nu$. This symmetry will act over pairs of flavors for each mixing angle $\theta_{ij}$. Then for the electroweak theory, the appropriate symmetry group is $SU(3)_\nu\otimes SU(2)_L\otimes U(1)$. For the $SU(3)_\nu$ symmetry, we have to be careful because this symmetry corresponds to three different matrices obeying the $SU(2)$ symmetry, but each one operating over two of the three available spaces. Then the $SU(3)_\nu$ representation in this case, can be expressed through matrices acting over two spaces. If we interpret the matrices $SU(2)_\nu$ as the Yang Baxter $R$-matrices, then we can express them as

\begin{eqnarray}
R_{1, 2}e_1\otimes e_2\otimes e_3=(R_{1, 2}e_1\otimes e_2)\otimes\hat{I}_{1\times1}e_3,\nonumber\\
R_{2, 3}e_1\otimes e_2\otimes e_3=\hat{I}_{1\times1}e_1\otimes(R_{2, 3} e_2\otimes e_3),\nonumber\\
R_{1, 3}e_1\otimes e_2\otimes e_3=R_{1, 3}e_1\otimes (\hat{I}_{1\times1}e_2)\otimes e_3.
\end{eqnarray}
In each case, each matrix only acts over two of the three spaces available. In this case, each space corresponds to one specific mass eigenvalue. We can still invert the matrices correspondingly, making them matrices acting on flavor spaces instead of eigenvalue spaces. In general, the $SU(2)_{\nu_{ij}}$ matrices acting over a two-dimensional space, can be expanded in terms of the Pauli matrices as

\begin{equation}
\sigma_1=
\begin{pmatrix}
0 & 1 \\
1 & 0 
\end{pmatrix},\;\;\;\sigma_2=
\begin{pmatrix}
0 & -i \\
i & 0 
\end{pmatrix},\;\;\;\sigma_3=
\begin{pmatrix}
1 & 0 \\
0 & -1 
\end{pmatrix}.    
\end{equation}
In this way, we can expand the following expression

\begin{equation}   \label{Upa}
U(\theta_{ij})=cos\frac{\theta_{ij}}{2}-i{\bf\sigma}\cdot{\bf n}sin\frac{\theta_{ij}}{2}.
\end{equation}
We can then construct the matrices related to the mixing angles $\theta_{12}$, $\theta_{13}$ and $\theta_{23}$. The matrices are 

\begin{eqnarray}   \label{mamamia}
\begin{pmatrix}
m_2 \\
m_3 
\end{pmatrix}=
\begin{pmatrix}
cos\frac{\theta_{23}}{2} & isin\frac{\theta_{23}}{2} \\
isin\frac{\theta_{23}}{2} & cos\frac{\theta_{23}}{2}
\end{pmatrix}\begin{pmatrix}
m_\mu \\
m_\tau 
\end{pmatrix}
,\;\;\;\begin{pmatrix}
m_1 \\
m_3 
\end{pmatrix}=
\begin{pmatrix}
cos\frac{\theta_{13}}{2} & sin\frac{\theta_{13}}{2} \\
-sin\frac{\theta_{13}}{2} & cos\frac{\theta_{13}}{2}
\end{pmatrix}\begin{pmatrix}
m_e \\
m_\tau
\end{pmatrix},\;\;\; \nonumber\\
\begin{pmatrix}
m_1 \\
m_2 
\end{pmatrix}=
\begin{pmatrix}
e^{i\frac{\theta_{12}}{2}} & 0 \\
0 & e^{-i\frac{\theta_{12}}{2}}
\end{pmatrix}\begin{pmatrix}
m_e \\
m_\mu 
\end{pmatrix}   
\end{eqnarray}
The expressions (\ref{mamamia}) were derived from eq. (\ref{Upa}) after considering the inverse of $U(\theta_{ij})$ as $U(-\theta_{ij})=U^{-1}(\theta_{ij})$, in order to solve for the variables $m_1$, $m_2$ and $m_3$. It is not difficult to notice that $R_{i,j}=U(-\theta_{ij})$ in this scenario. Finally, it is possible to see that the action of the matrices acting over two among three different spaces, can be represented as follows

\begin{eqnarray}   \label{mamamia1}
\begin{pmatrix}
m_1\\
m_2 \\
m_3 
\end{pmatrix}=
\begin{pmatrix}
1 & 0 & 0\\
0 & \;\;\;SU(2)_{\nu_{2, 3}}\\
 0& 
\end{pmatrix}\begin{pmatrix}
m_e \\
m_\mu \\
m_\tau 
\end{pmatrix}
,\;\;\;\begin{pmatrix}
m_1 \\
m_2 \\
m_3 
\end{pmatrix}=
\begin{pmatrix}
 SU(2)_{\nu_{1}} & 0\\
0 & 1 & 0 \\
0 & SU(2)_{\nu_{3}}
\end{pmatrix}\begin{pmatrix}
m_e \\
m_\mu \\
m_\tau 
\end{pmatrix},\;\;\; \nonumber\\
\begin{pmatrix}
m_1 \\
m_2 \\
m_3
\end{pmatrix}=
\begin{pmatrix}
e^{i\frac{\theta_{12}}{2}} & 0 & 0\\
0 & e^{-i\frac{\theta_{12}}{2}} & 0\\
0 & 0 & 1
\end{pmatrix}\begin{pmatrix}
m_e \\
m_\mu \\
m_\tau 
\end{pmatrix}   
\end{eqnarray}
In the last expression, the $SU(2)_{\nu_{1,2}}$ matrix is defined as

\begin{equation}   \label{mamamia2}
SU(2)_{\nu_{1, 2}}=\begin{pmatrix}
e^{i\frac{\theta_{12}}{2}} & 0 \\
0 & e^{-i\frac{\theta_{12}}{2}}
\end{pmatrix}
\end{equation}
The order of the operations over the flavor vector is 

\begin{equation}   \label{mamamiaa}
\begin{pmatrix}
m_1\\
m_2 \\
m_3 
\end{pmatrix}=U(\theta_{1, 2})U(\theta_{\theta_{1, 3}})U(\theta_{2, 3})\times\begin{pmatrix}
m_e\\
m_\mu \\
m_\tau
\end{pmatrix}    
\end{equation}
Inside this last expression, it is understood that each matrix $U(\theta_{\nu_{i, j}})=R_{i, j}$ acts over two of the three available spaces. 

\subsection{The electroweak theory: Extensions including flavor mixing}

\begin{figure}[h!]
\centering
\includegraphics[scale=0.7]{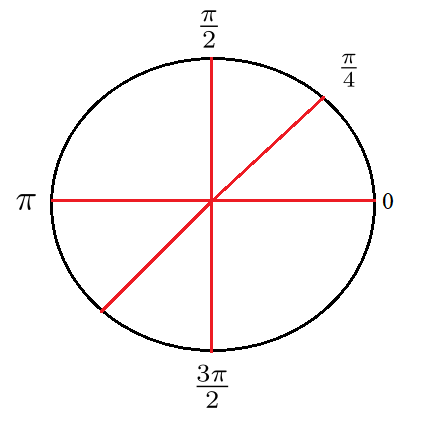} 
\caption{Circle illustrating the key values for the angle $2\theta_{12}$. The angles $0$, $n\pi$ (n integer), correspond to the same state up to a total sign. The same applies to any pair of angles connected by any line crossing the center of the circle. The angle $2\theta_{12}=72 deg$   \label{fig3}}   
\end{figure}

\begin{figure}[h!]
\centering
\includegraphics[scale=0.7]{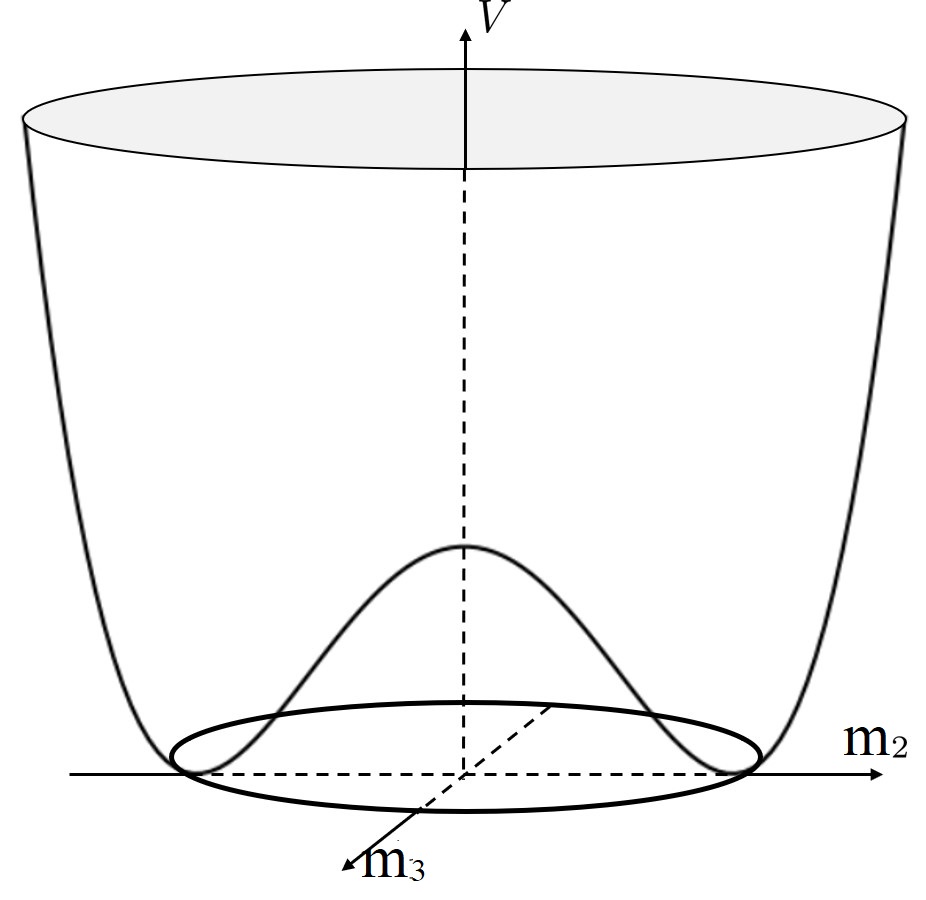} 
\caption{Illustration of the neutrino oscillation process as a jumping process between three different degenerate vacuum states. The neutrino takes the mass $m_3$ for some particular vacuum state and $m_2$ for another one. For simplicity we omit the state corresponding to $m_1$. The same illustration applies for the flavor oscillation. \label{fig2}}
\end{figure}

The electroweak theory operates under the gauge symmetry $SU_L(2)\otimes U(1)$, which is spontaneously broken, following the pattern $SU_L(2)\otimes U(1)\to U(1)_{em}$. As can be seen, the $SU_L(2)$ gauge symmetry pattern only involves left-handed particles, understanding that the neutrino itself is left-handed. Still, due to the neutrino oscillations, there is another symmetry related to this phenomenon. In order to involve the three flavors and eigenvalues, we could think over the $SU_{fl}(3)$-symmetry involving the three flavors and/or eigenvalues $m_1$, $m_2$ and $m_3$. However, in order to simplify matters, in this paper we will intend to make a simplification by using as first approximation $m_1\approx m_2$ as the observations suggest. Under this initial approximation, $\theta_{13}\to0$ and $\theta_{12}\approx\theta_{23}$. Although this case is not the real one, still it will help us to make a standard formulation which we can improve in future papers. We will then take an approximate $SU_{fl}(2)$ (flavor) symmetry as the appropriate one, extending then the usual $SU_L(2)\otimes U(1)$ symmetry of the electroweak theory to the symmetry $SU_{fl}(2)\otimes SU(2)_L\otimes U(1)$. More exactly, the exact expression for the symmetry breaking pattern is then

\begin{equation}
SU_{fl}(3)\otimes SU(2)_L\otimes U(1)\approx SU_{fl}(2)\otimes SU(2)_L\otimes U(1)\to SU_{fl}(2)\otimes U_{em}(1).     
\end{equation}
This means that the flavor symmetry is also unbroken, besides the $U(1)$-symmetry corresponding to the electromagnetism. Under the present simplification, we can reduce the result (\ref{mamamiaa}) to

\begin{equation}   \label{mamamiaa2}
\begin{pmatrix}
m_1\\
m_2 \\
m_3 
\end{pmatrix}\approx \begin{pmatrix}
m_2 \\
m_3 
\end{pmatrix}=U(\theta_{2, 3})\begin{pmatrix}
m_{f1} \\
m_{f2}
\end{pmatrix}=\begin{pmatrix}
cos\frac{\theta_{23}}{2} & isin\frac{\theta_{23}}{2} \\
isin\frac{\theta_{23}}{2} & cos\frac{\theta_{23}}{2}
\end{pmatrix}\begin{pmatrix}
m_{f1} \\
m_{f2}
\end{pmatrix},
\end{equation}
where we have taken $m_2\approx m_1$ and we have also reduced the analysis to two flavors just for illustrating the method. Since there are only two vacuum states in agreement with eq. (\ref{mamamiaa2}), the charge corresponding to symmetry involved in the flavor mixing process, maps the neutrino from one flavor to another. Then two applications of the same charge over the vacuum state, will bring the quantum state to the original flavor. This means that if we define the broken generator as $Q_a$ and then apply it to the ground state, then two successive applications gives us the identity as

\begin{equation}
Q^2_a\vert m_{f1}>=\vert m_{f1}>.    
\end{equation}
Additionally, we have the condition

\begin{equation}
Q_a\vert m_{f1}>=\vert m_{f2}>.    
\end{equation}
In this way, we have the following result for the broken charge $U=e^{-iQ_a\theta_{23}}$, with $2\theta_{23}=\frac{\pi}{2}$ to mix the flavors, in agreement with the triangles defined in the figure (\ref{fig1}). The symmetry generator $Q_a$ in this case, would take a similar form to the corresponding Pauli operator $Q_a=\frac{\sigma_a}{2}$. The present formulation suggests a direct mix between the flavors when $2\theta_{23}\approx \frac{\pi}{2}$ and a return to the original $2\theta_{23}=\pi$ for the flavors (up to a total sign). This means that if the angle of rotation is $\theta_{23}=\frac{\pi}{4}$, we then get a perfect mix between both flavors. In the meantime, when the mixing angle is $\theta_{23}=\frac{\pi}{2}$, we recover the negative value of the original vacuum state, which is redundant with it. Then the angles $\theta_{23}=\frac{3\pi}{4}$ and $\theta_{23}=\pi$, correspond to the same states represented by the first two angles $\theta_{23}=\frac{\pi}{4}$ and $\theta_{23}=\frac{\pi}{2}$ (up to a change of sign). The figure (\ref{fig3}) illustrates this aspect. From figure (\ref{fig2}), we can understand the effect of $Q_a$ on the ground state $m_2$ as a mapping of this state to the state corresponding to the mass $m_3$. This is the case because the potentially broken charge $Q_a$, which is actually never broken in reality, just shifts the degenerate vacuums and maps them one over the other. 

\section{Triangular relation between mixing angles and masses}   \label{YANG}

The triangular relations are the simplest possible relations among the three mixing angles. However, this is a simplification which will be enough for the purpose of calculations, understanding that the deviations from the triangular geometry will bring uncertainty into the calculations. The simplification is not accurate because the sum of the double of the mixing angles is not necessarily $180deg$. Then in general we have

\begin{equation}   \label{Superangle}
2\theta_{12}+2\theta_{13}+2\theta_{23}\neq180deg.    
\end{equation}
Yet still we will use the approximation $2\theta_{12}+2\theta_{13}+2\theta_{23}\approx180deg$, understanding that possible deviations from the triangle condition might come from possible $CP$ violations and other physical aspects which we are ignoring in this calculation, such as quantum corrections. The starting point is the standard definition of a vacuum state defined in eq. (\ref{standardvac}). In the same way as we can define vacuum states for each flavor, we can also define vacuum states with respect to the eigenvalue index $i$ for the masses $m_i$, with $i=1, 2, 3$. The triangular relations proposed in this paper, are generated with respect to the basis of the mass eigenvalues $m_i$. We could perfectly generate the triangles with respect to the flavor basis, but this is not the simplest way to work into the problem. Then instead of imagining a neutrino moving through spacetime and oscillating between different flavors, we will take the equivalent approach suggesting that the neutrinos with some specific flavor, have an oscillatory mass between three different values $m_i$. This point of view is allowed by the mixing matrix (\ref{matrix}) if we calculate its inverse.
\begin{figure}[h!]
\centering
\includegraphics[scale=0.7]{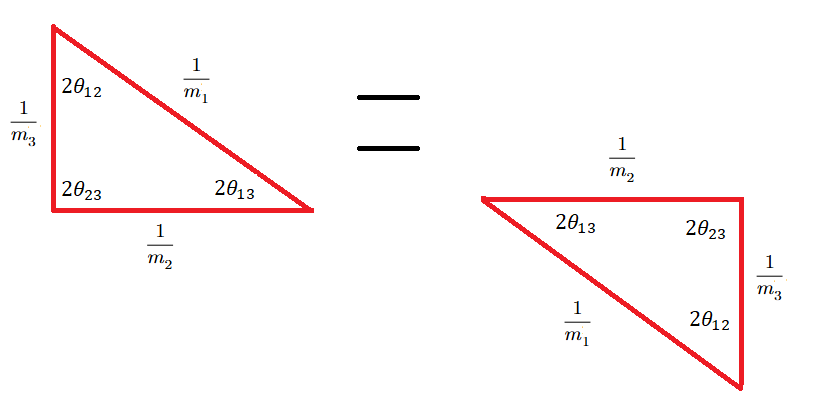} 
\caption{Triangular relation between the mass eigenvalues in agreement with the QYBE. For the case of normal hierarchy, each side of the triangles represents the inverse of the mass for each eigenvalue $1/m_i$. The mixing angles are illustrated, with $\theta_{23}\approx\pi/2$. Interestingly, if $\theta_{13}\to0$, then $m_1\approx m_2$. For illustration, we have taken $\theta_{23}\approx 45 deg$ \label{fig1}}
\end{figure}
In the figure (\ref{fig1}), we illustrate the triangles representing the relation between the mixing angles and the mass eigenvalues. The two triangles illustrated are indistinguishable between each other. In this case we graphically impose the Quantum Yang Baxter Equations (QYBE), understanding that the direction of motion of the neutrinos can change. Additionally, the equivalence between the triangles shows that the signature of the angles is not so important and instead, their absolute value is what really matters.
Although we input the QYBE in this scenario, at this point they are not really necessary for analyzing the relations between the different mass eigenvalues and the mixing angles. The QYBE guarantee that we have to apply the superposition of all the possible ground (flavor) states \cite{M1, M2, M3}. The neutrino experiments suggest two possibilities for the values of masses, they are: 1). Normal ordering (NO), where $m_3>>m_1\approx m_2$. 2). Inverted Ordering (IO), where $m_1\approx m_2>>m_3$. We will start by analyzing the first possibility.

\subsection{Normal Ordering with the triangular relations}

As we have mentioned before, the neutrino never breaks the symmetry of the vacuum spontaneously and then it is unable to select some specific mass (ground state). Instead, it keeps oscillating between the three vacuum states without selecting anyone in particular. The neutrino, identified by its flavor, is our order parameter. It is for this reason that the sum of the possible flavors vanish. In this scenario, we will take each eigenvalue $m_1$, $m_2$ and $m_3$ to be represented by a line forming a triangle. If we want this triangular model to obey the Normal Ordering, then each triangle line would correspond to the inverse of the eigenvalue mass $1/m_i$. Then for example, the side with label $1$ in the figure (\ref{fig1}) is equivalent to the value $1/m_1$. The same applies for the other sides. The triangle is taken approximately as a rectangle triangle, suggesting then $\vert\theta_{23}\vert\approx\pi/2$. Although our value derived in the previous section with $\vert\theta_{23}\vert\approx 39deg$. For the moment, we will tolerate this discrepancy, understanding that the triangle is just a simplification of the way how the neutrino behaves. Although both values, $45deg$ and $39deg$ for $\theta_{23}$ are inside the experimental uncertainties, the value of $39deg$ is not generally consistent with the present best-fit regions \cite{Bestfit, Bestfit2}. Still, as an initial approximation, for the purpose of this paper, it is still correct to take the triangle approximation as an initial ansatz. The Pythagoras theorem for the triangle, under the present approximation, suggests that

\begin{equation}   \label{Pita}
\frac{1}{m_1^2}=\frac{1}{m_2^2}+\frac{1}{m_3^2}.
\end{equation}
Then if $m_3>>m_2>m_1$, then the relation $m_2\approx m_1$ emerges naturally. This means that $m_2$ and $m_1$ have values of the same order of magnitude with a normal hierarchy. It is interesting to notice that when $\theta_{13}\to0$, then $m_1=m_2$ under the triangular formulation. This means that the proximity between the values $m_1$ and $m_2$ is explained by the fact that the angle $\theta_{13}$ is very small. Then we can interpret $\theta_{13}$ as the angle mixing $m_1$ and $m_2$. For proving this statement, we can use the mixing matrix (\ref{matrix}) and expand explicitly the relations between vacuums as in eq. (\ref{linear_comb}). If we express the results in terms of the neutrino masses and if we set $\theta_{13}=0$ in this equation and if additionally we work with the inverse of the matrix, we get

\begin{eqnarray} \label{linear_comb2}
m_1 & = & c_{12}m_e-s_{12}c_{23}m_\mu+s_{12}s_{23}m_\tau,\nonumber\\
m_2 & = & s_{12}m_e + c_{12}c_{23}m_\mu-c_{12}s_{23}m_\tau, \nonumber \\
m_3 & = & c_{13}s_{23}m_\mu+c_{13}c_{23}m_\tau.
\end{eqnarray}
It does not matter if we express the oscillation in terms of the vacuum states or in terms of the masses because the results will be equivalent anyway. If the triangular relations were taken seriously, then $m_1=m_2$ when $\theta_{13}=0$, in agreement with the triangle defined in the figure (\ref{fig1}). This means that the angle $\theta_{13}$ would be the angle mixing $m_1$ and $m_2$. Additionally, from eq. (\ref{linear_comb2}) we would have $\theta_{12}=\theta_{23}=45deg$, which is the only possibility for the case $\theta_{13}$ if we want to keep the condition $2\theta_{12}+2\theta_{13}+2\theta_{23}=180deg$. Additionally, from eq. (\ref{linear_comb2}), if $m_1=m_2$, we would get $m_\mu=m_\tau$. These results are not accurate but they indicate how the triangle relation would behave in these extreme cases. This would also fix the value $m_3=\frac{1}{2}m_\mu=\frac{1}{2}m_\tau$. Here again we remark that these results are not real and instead, they would correspond to the hypothetical extreme case where $\theta_{13}$. Later we will analyze in deep detail these statements, showing the conditions under which the triangular relations would be valid. For the moment, let's try to derive the mass predictions done through the triangle relations. The experimental evidence suggests that \cite{Exp}

\begin{equation}   \label{masa1}
m_2^2-m_1^2\approx0.759\times10^{-4}eV^2.    
\end{equation}
\begin{equation}   \label{masa2}
m_3^2-m_1^2\approx m_3^2-m_2^2\approx23.2\times10^{-4}eV^2.    
\end{equation}
Here again, we perceive $m_3>>m_2\approx m_1$, which is consistent with the triangle shown in the figure (\ref{fig1}). From the triangles on the figure (\ref{fig1}), we can obtain the mixing angles as a function of the mass ratio of the different neutrino masses. We then have the following relations

\begin{equation}   \label{mici}
sin 2\theta_{13}\approx\frac{m_1}{m_3}, \;\;\;sin 2\theta_{12}\approx\frac{m_1}{m_2},\;\;\;sin 2\theta_{23}\approx 1.    
\end{equation}
This result, represents important constraints between the mixing angles and the values taken by the neutrino masses. If we combine the results obtained in eq. (\ref{masa1}) and (\ref{masa2}), with the relations (\ref{mici}), we can the obtain the explicit values for each mass $m_i$, as long as we are able to measure the mixing angles. The following relations emerge

\begin{eqnarray}   \label{squaremasses}
m_2^2-m_1^2=m_2^2(1-sin^2(2\theta_{12}))\approx0.759\times10^{-4}eV^2,\nonumber\\
m_3^2-m_1^2=m_3^2(1-sin^2(2\theta_{13}))\approx23.2\times10^{-4}eV^2.\;\;
\end{eqnarray}
Note that for keeping the condition $m_3>>m_1$, we need to satisfy $\theta_{13}$ to be small enough. This can explain why the experiments suggest that under normal ordering, $m_3$ would be much larger than the other values of masses. The same results would also suggest that $\theta_{12}$ has to be close to $45deg$ or at least on a value near to this. Here we can input our predictions of the previous section, with $\theta_{13}\approx 8.47deg$, which was the only input value from the previous section in order to derive the other mixing angles plus $\theta_{12}\approx 54deg$, which was a derived value. If we replace these values in eq. (\ref{squaremasses}), then we get

\begin{eqnarray}   \label{masspred}
m_3\approx 5.04\times10^{-2}eV, \;\;\;m_2\approx2.82\times10^{-2}eV,\;\;\;
m_1\approx 2.68\times10^{-2}eV.
\end{eqnarray}
\subsection{Inverted ordering}

Here we will demonstrate that from the perspective of a triangular formulation, the Inverted Ordering brings out some inconsistencies. For inverted ordering, each side of the triangles in the figure (\ref{fig1}) would correspond to the values of mass $m_1$, $m_2$ and $m_3$ and not to their inverse. Then the relation (\ref{Pita}) would change to 

\begin{equation}
m_1^2=m_2^2+m_3^2.    
\end{equation}
In this case we still have $m_1\approx m_2>>m_3$. The trigonometric relations (\ref{mici}) would be modified to their corresponding inverses

\begin{equation}   \label{mici2}
sin 2\theta_{13}\approx\frac{m_3}{m_1}, \;\;\;sin 2\theta_{12}\approx\frac{m_2}{m_1},\;\;\;sin 2\theta_{23}\approx 1.    
\end{equation}
In this situation, we can still use the same mixing angles used in eqns. (\ref{masa1}) and (\ref{masa2}). However, it is not hard to realize that for this case, we would obtain complex values for the masses. Then we rule out this possibility. The triangular formulation proposed here, was formulated in 2024 by the same author in \cite{Triangle1, Triangle12}. By that time however, the interpretation of the flavor as an order parameter of the system was not proposed nor the analysis of the problem from the perspective of symmetry. Another reference where it was suggested a similar triangular formulation, came out in 2025 in \cite{Triangle2}. Although the author in \cite{Triangle2} used the same superposition condition proposed after eq. (\ref{Superangle}), in such a case each vertex of the triangle was represented by $m_i^2$, while in \cite{Triangle1, Triangle12}, as well as in this paper each side of the triangle is equivalent to $\frac{1}{m_i^2}$, which is a key ingredient for excluding the inverted ordering (IO) as this section has showed.

\subsection{Invariance of $\theta_{13}$ and the exchange $\theta_{12}\to\theta_{23}$}

The triangular formulation is a good approximation for predicting the mass values $m_1$, $m_2$ and $m_3$. However, it should not be taken literally as the absolute truth for the analysis of the neutrino oscillation pattern. In the next section we will make new predictions on the values of mass by using a more direct approach based on the arguments of the section (\ref{Vacuumpers}). The general argument of this paper is to take the neutrino flavor as an order parameter. This implies the existence of three different vacuum states corresponding to each neutrino flavor. This was the key ingredient for deriving the expression (\ref{SumofVacumms}). If we calculate the inverse of the matrix (\ref{matrix}) and then develop the corresponding equations, we will get the following result 

\begin{eqnarray} \label{linear_comb3}
\vert0 \rangle_1 & = & c_{12}c_{13}\vert0\rangle\rangle_e-(s_{12}c_{23}+c_{12}s_{23}s_{13})\vert0\rangle\rangle_\mu+(s_{12}s_{23}-c_{12}c_{23}s_{13})\vert0\rangle\rangle_\tau, \nonumber \\
\vert0 \rangle_2 & = & s_{12}c_{13}\vert0\rangle_e +(c_{12}c_{23}-s_{12}s_{23}s_{13})\vert0\rangle_\mu-(c_{12}s_{23}+s_{12}c_{23}s_{13}))\vert0\rangle_\tau, \nonumber \\
\vert0 \rangle_3 & = & s_{13}\vert0\rangle_e + s_{23}c_{13}\vert0\rangle_\mu+c_{23}c_{13}\vert0\rangle_\tau.
\end{eqnarray}
If we interpret the result among the three possible vacuums, and then take again the neutrino, but this time with the labels $1, 2, 3$, as the order parameter, then we get

\begin{equation}   \label{SumofVacumms2}
\vert0 \rangle_1+\vert0 \rangle_2+\vert0 \rangle_3=0,    
\end{equation}
which resembles the flavor version of this expression defined in eq. (\ref{SumofVacumms}). If we combine the results (\ref{linear_comb3}) and (\ref{SumofVacumms2}), we get

\begin{eqnarray}   \label{ImportantConditions3}
s_{13}+s_{12}c_{13}+c_{12}c_{13}=0,\nonumber\\
s_{23}s_{13}-s_{12}c_{23}-c_{12}s_{23}s_{13}+c_{12}c_{23}-s_{12}s_{23}s_{13}=0\nonumber\\
c_{23}c_{13}+s_{12}s_{23}-s_{12}c_{23}s_{13}-c_{12}s_{23}-c_{12}c_{23}s_{13}=0,
\end{eqnarray}
which helps us to find some constraints between the mixing angles. In this alternative scenario, the roles of the angles $\theta_{12}$ and $\theta_{23}$ are exchanged in the first equation in (\ref{ImportantConditions3}). In this way we will obtain $\theta_{12}\approx 39deg$. When we solved the angles on the previous case, without loss of generality, we mapped the angles as $\theta_{ij}\to-\theta_{ij}$, changing then the frame of reference. This can be seen from the solutions obtained from eq. (\ref{ImportantConditions}) and (\ref{ImportantConditions2}). This also explains the change in the signature of some of the terms under analysis. If we repeat the same arguments for eqns. (\ref{ImportantConditions3}), then we get

\begin{eqnarray}   \label{ImportantConditions4}
-s_{13}-s_{12}c_{13}+c_{12}c_{13}=0,\nonumber\\
s_{23}s_{13}+s_{12}c_{23}-c_{12}s_{23}s_{13}+c_{12}c_{23}+s_{12}s_{23}s_{13}=0\nonumber\\
c_{23}c_{13}+s_{12}s_{23}-s_{12}c_{23}s_{13}+c_{12}s_{23}+c_{12}c_{23}s_{13}=0.
\end{eqnarray}
The first equation in this expression can be simplified to be

\begin{equation}
tan\theta_{13}=cos\theta_{12}-sin\theta_{12}.
\end{equation}
This result is analog to the one obtained in eq. (\ref{1st1}), but with the result solved for $\theta_{12}\approx 39deg$, obtained when we input $\theta_{13}=8.47deg$. Now we can proceed to solve $\theta_{23}$ for the redundant equations defined by the second and third equations inside (\ref{ImportantConditions4}). By entering the results, we get $\theta_{23}\approx 54deg$. This means that the roles of $\theta_{12}$ and $\theta_{23}$ are exchanged when we change the vacuum basis. This also means that the only invariant angle under this change of basis is the mixing angle $\theta_{13}$. Finally, here we also remark that the observations suggest that the correct result for the angle $\theta_{23}$ would be $\theta_{23}\approx 36deg$, which is just the complementary angle to $\theta_{23}\approx 54deg$. Here again, under the assumption that we are only considering $\vert\theta_{ij}\vert$ for all the angles, or equivalently, the angle and its complementary value are equivalent. This guarantees that the probability remains the same after changing the signature of the angles \cite{parameters}.

\section{Future analysis and predictions}   \label{Predictionsanalysis}

This paper did not provide a mechanism for the generation of mass of the neutrinos. Yet still it gives some clues about this aspect by considering the flavor symmetry as the key ingredient in the analysis since we took the flavor as the order parameter of the system. The proposed formulation predicts certain values for the mixing angles as well as for the mass eigenvalues. In future analysis, we will be studying the flavor symmetry deeply. In particular, we will be exploring deeply in detail the symmetry breaking pattern $SU_{fl}(3)\otimes SU(2)_L\otimes U(1)\approx SU_{fl}(2)\otimes SU(2)_L\otimes U(1)\to SU_{fl}(2)\otimes U_{em}(1)$. The essential idea is that the flavor symmetry is not spontaneously broken. We suspect that the mechanism of generation of the neutrino masses is connected with this symmetry oscillation but this aspect will be analyzed in coming papers. Future analysis also requires a full conciliation between the triangular formulation and the symmetry approach, which is fundamental. In summary, we have predicted the following values for the mixing angles 

\begin{equation}
\theta_{12}\approx 54 deg\;\;or -36deg,\;\;\;\theta_{13}\approx 8.47 deg,\;\;\;\theta_{23}\approx 39deg.    
\end{equation}
In this paper, we have to specify that we are taking the angles and their complementary as equivalent. In this way, we will consider the absolute values of the angles, namely $\theta_{12}\approx 54$ will be equivalent to $\theta_{12}\approx 36deg$, considering $\vert\theta_{12}\vert$ as the relevant quantity. This means that we want to keep all the angles under the condition $2\theta_{ij}\leq \pi/2$ ar equivalently, $2\theta_{ij}\leq 90deg$. Note that if we consider absolute values of the angles, as the QYBE suggests, then $\theta_{12}\approx 36deg$ is also a predicted value inside this formulation. Here we remark once again that $\theta_{13}$ is an invariant under exchange of basis between the flavor basis and the eigenvalue basis, while the other two mixing angles exchange roles during such exchanges. Finally, here we predicted the values of masses given in eq. (\ref{masspred}). Still, in order to guarantee accurate values, we will consider potential CP violations in future papers. This in light of the discrepancies between DUNE and T2K \cite{T2K, NOVA, DUNE}.

\subsection{Expected experiments verifying the theoretical arguments}

There are currently major experiments, namely, Nova and T2K \cite{T2K, NOVA, DUNE}. It is well-known that the probability of transition from one flavor to another, for example $\nu_\mu\to\nu_e$ is a function of $\vert\Delta m_{ij}^2\vert$, namely, the neutrino mass. In addition, the probability of transition also depends on the mixing angles \cite{T2K}. Since in this paper we have demonstrated that the triangular formulation only allows Normal Ordering (NO), this will generate constraints on the possible values on the mentioned oscillation probability. In this paper we do not intend to involve the potential CP violations but instead, we ignore it for the moment, still understanding that the might affect the transition probabilities too. In future papers we intend to include the effects of this additional parameter in our formulation. Our mass predictions depend strongly on the mixing angles obtained. The QYBE suggests that the signature of the angles $\theta_{ij}$ is irrelevant and instead, the absolute value of the angles is what matters. This can be seen from the figure (\ref{fig1}). This means that the angle $\theta_{12}\approx 54deg$ is just equivalent to $\theta_{12}\approx 36deg$ if we consider the absolute value of the complementary angle. Both angles reproduce the same value of $sin^2(2\theta_{12})$ and that's why we claim consistency with the experiments in this paper. If the experiments in future find some CP violation, then this would certainly change a few results in this paper in the sense that the values of some of the mixing angles will change. In fact, the CP violations would just bring out small corrections to the triangular formulation. The fact that the sum of the double of the angles does not close to $180deg$, means that there is some hidden physics which we are not considering yet. This physics can be hidden inside the experimental results. The author intends to follow the coming results from T2K and Nova. Finally we are currently analyzing the joint measurement done by Nova and T2K, \cite{T2K}. These experiments have serious discrepancies on the values obtained in relation to the CP violation. The refinement of the future experiments will determine if there is really CP violation. Further advances on the experiments will bring us the opportunity of testing our model, in particular, the fact that the neutrino flavor is the order parameter can be tested if the uncertainty on the values of the mixing angles is reduced inside all the experiments. In other words, the coming experiments will help us to improve our model or to exclude it totally. Finally, in future papers we will work on the construction of the Lagrangian of the theory and on the mechanism of generation of the neutrino masses.

\section{Conclusions}   \label{concl}

In this paper we have formulated the mixing angles of the neutrino oscillation phenomena by understanding this oscillation as a consequence of a vacuum degeneracy. Then the neutrino oscillates between three vacuum flavors but the external perturbations are not strong enough for forcing the neutrino to select some specific vacuum state. Then the symmetry responsible for this oscillation is not broken spontaneously. The proof of this statement is in the fact that if we take the vacuum flavor as the order parameter of the system, we are then able to reproduce the observed mixing angles. This is possible because it is well-known that the superposition of all the possible values taken by the order parameter is zero. From this standard result, we obtain constraints for the mixing angles. Subsequently, by using a triangular formulation, consistent with the Quantum Yang Baxter Equations (QYBE), we could make the predictions for the values of mass. The connection between the QYBE and the vacuum degeneracy, were understood in \cite{M1,M2,M3}. By using the mentioned constraints, we were able to find the exact values for the three neutrino masses $m_1$, $m_2$, and $m_3$, in a similar fashion as it has been proposed in \cite{Triangle1, Triangle12}. The same constraints suggest that the neutrinos must follow a normal hierarchy on their mass values. From the perspective of this paper, the electroweak theory in its standard form, cannot give mass to the neutrinos because the neutrinos never break the symmetry of the ground state spontaneously. The proposed approach assumes no violation of the CP-symmetry as an initial approximation. We also did an analysis of the final symmetry pattern associated with the electroweak theory, but considering the simplification $m_1\approx m_2$, which allows us to take the flavor symmetry as $SU(2)_{fl}$. The flavor symmetry is not spontaneously broken by the neutrino in this case. The symmetry arguments explain the emergence of the values taken by the neutrino mixing angles. The same arguments also explain why the mixing angle $\theta_{13}$ is so small. Finally, we could prove that a change of the flavor basis into the eigenvalue basis, generates exchange on the roles of the angles $\theta_{23}$ and $\theta_{12}$, while the angle $\theta_{13}$ is fixed under the change of basis. For all the mixing angles we have taken their absolute value, namely, $\vert\theta_{ij}\vert$; particularly when we use the condition of taking the flavor as the order parameter. This also means that we are taking the mixing angles and their complements are taken as the same. This guarantees that the transition probability between flavors is not affected after changing the signature of the mixing angles \cite{parameters}.



\begin{thebibliography}{0}

\bibitem{the_neutrino}
H. Bethe and R. Peierls, {\it The "Neutrino"}, Nature, {\bf 133}, 532 (1934).

\bibitem{pdg}
(Particle Data Group), P. A. Zyla, et. al, {\it 2020 Review of Particle Physics}, Review of Particle Physics, Progress of Theoretical and Experimental Physics, Volume 2020, Issue 8, August 2020, 083C01.

\bibitem{glashow}
S. L. Glashow, {\it The renormalizability of vector meson interactions}, Nucl. Phys. 10, 107, 1959.

\bibitem{salam}
A. Salam, {\it Weak and Electromagnetic Interactions}, Conf.Proc.C 680519 (1968) 367-377.

\bibitem{weinberg}
S. Weinberg, {\it A Model of Leptons}, Phys. Rev. Lett. 19, 1264.

\bibitem{zeromass}
Peskin M. E and Schroeder D. V.; {\it An introduction to Quantum Field Theory}, CRC press, Taylor and Francis
group, 6000Broken Sound Parkway, NW Suite 300, Boca
Raton Fl. 33487-2742, (2018).

\bibitem{weyl}
M. Drewes, {\it The Phenomenology of Right Handed Neutrinos}, Int. J. Mod. Phys. E, {\bf 22}, 1330019 (2013).

\bibitem{sk}
Y. Fukuda et at. (Super-Kamiokande Collaboration), {\it Evidence for Oscillation of Atmospheric Neutrinos}, Phys. Rev. Lett. {\bf 81}, 1562, 1998.

\bibitem{sno}
Q. R. Ahmad, et al. (SNO Collaboration), Measurement of the rate of $\nu_e$ + d $\to$ p + p + $e^-$ interactions produced by 8B Solar neutrinos at the Sudbury Neutrino Observatory, Phys. Rev. Lett. {\bf 87}, 071301, 2001.

\bibitem{pontecorvo}
B. Pontecorvo, {\it Neutrino Experiments and the Problem of Conservation of Leptonic Charge}, Sov.Phys.JETP 26 (1968) 984-988, Zh.Eksp.Teor.Fiz. 53 (1967) 1717-1725.

\bibitem{mns}
Z. Maki, M. Nakagawa and S. Sakata, {\it Remarks on the Unified Model of Elementary Particles}, Prog.Theor.Phys. 28 (1962) 870-880.

\bibitem{parameters}
P. A. Zyla, et. al (Particle Data group), {\it Neutrino mixing}, Prog. Theor. Exp. Phys. 2020, 083C01, (2020). 

\bibitem{Seesaw}
A. Batra, P. Bharadwaj, S. Mandal, R. Srivastava and J. W. F. Valle, {\it Phenomenology of the simplest linear seesaw mechanism},  J. High Energ. Phys. 2023, 221 (2023).

\bibitem{Seesaw2}
T. Yanagida, {\it Horizontal Symmetry and Masses of Neutrinos}, Prog. Theor. Phys. Vol. 64, No. 3, 1980.

\bibitem{Delta1}
G. Altarelli and F. Feruglio, {\it Discrete flavor symmetries and models of neutrino mixing}, Rev. Mod. Phys. {\bf 82}, 2010.

\bibitem{Delta2}
E. Ma, {\it Dark scalar doublets and neutrino tribimaximal mixing from symmetry}, Phys.Lett. B671:366-368,2009.

\bibitem{Delta3}
V. Puyam, S. Robertson Singh and N. Nimai Singh, {\it Deviation from Tribimaximal mixing using $A_4$ flavour model with five extra scalars}, Nucl. Phys. B, 983, 2022, 115932.

\bibitem{Delta4}
E. Ma, {\it Tribimaximal neutrino mixing from a supersymmetric model with $A_4$ family symmetry}, Phys. Rev. D 73, 057304, 2006.






\bibitem{Triangle1}
I. Arraut, {\it The neutrino mass: A triangular formulation}, 	arXiv:2406.02641 [physics.gen-ph].

\bibitem{Triangle12}
I. Arraut and E. Arrieta-Diaz, {\it Neutrino oscillations from the perspective of the quantum Yang-Baxter equations}, 	arXiv:2409.00560 [hep-ph]. 

\bibitem{Triangle2}
M. F. H. Seikh, {\it A geometrical approach to neutrino oscillation parameters}, AIP Advances 15, 105204 (2025).

\bibitem{M1}
I. Arraut, {\it The origin of the mass of the Nambu–Goldstone bosons}, Int.J.Mod.Phys.A 33 (2018) 07, 1850041.

\bibitem{M2}
I. Arraut, {\it The Quantum Yang Baxter conditions: The fundamental relations behind the Nambu-Goldstone theorem}, Symmetry 11 (2019) 803.

\bibitem{M3}
I. Arraut, {\it The Nambu-Goldstone theorem in non-relativistic systems} , Int.J.Mod.Phys.A 32 (2017) 1750127.

\bibitem{TB}
H. Goldstein, {\it Classical Mechanics (2nd ed.)}, Reading, MA: Addison–Wesley, ISBN 978-0-201-02918-5, (1980).

\bibitem{Nambu2}
Nambu, Y., J.; {\it From Yukawa's Pion to Spontaneous Symmetry Breaking}, Phys. Soc. Jap. {\bf 76}, 111002 (2007).

\bibitem{Nambu3}
Nambu, Y. and Jona-Lasinio, G.; {\it Dynamical Model of Elementary Particles Based on an Analogy with Superconductivity. II}, Phys. Rev. {\bf 124}, 246(1961).

\bibitem{Nambu4}
Nambu, Y. and Jona-Lasinio, G.; {\it Dynamical Model of Elementary Particles Based on an Analogy with Superconductivity. I}, Phys. Rev. {\bf 122}, 345 (1961).

\bibitem{Klad}
 K. Grotz and H.V. Klapdor, {\it The Weak Interaction in Nuclear, Particle and Astrophysics}, CRC Press, 1990, ISBN-10: 0852743130; ISBN-13: 978-0852743133.

 \bibitem{Bestfit}
 M. Fodroci and T. Kitabayashi, {\it Modified TM2 for Reproducing All Best-Fit Values of Neutrino Mixing Angles}, arXiv:2511.15111 [hep-ph].

 \bibitem{Bestfit2}
 I. Esteban, M. C. G.-Garcia, M. Maltoni, I. M.-Soler, J. P. Pinheiro, and T. Schwetz, J.
High Energy Phys. 12, 2024 (216). NuFIT 6.0 (2024), www.nu-fit.org.

\bibitem{Homomorphism}
H. F. Jones, {\it Groups, Representations and Physics}, CRC Press, (1998), ISBN-10:0750305045; ISBN-13:978-0750305044. 

\bibitem{Exp}
K.A. Olive et al. (Particle Data Group), Chin. Phys. C, 38, 090001 (2014).

\bibitem{T2K}
The NOvA Collaboration, The T2K Collaboration, {\it Joint neutrino oscillation analysis from the T2K and NOvA experiments}, Nature 646, 818–824 (2025).

\bibitem{NOVA}
M. A. Acero, et al. {\it Improved measurement of neutrino oscillation parameters by the NOvA experiment}, Phys. Rev. D 106, 032004 (2022).

\bibitem{DUNE}
K. Abe, et al. {\it Measurements of neutrino oscillation parameters from the T2K experiment using $3.6\times 10^21$ protons on target} Eur. Phys. J. C 83, 782 (2023).


\end{thebibliography}

\end{document}